\documentclass[aps,
               pra,
               twocolumn,
               amsmath,
               amssymb,
               superscriptaddress,
               reprint,
               10pt
              ]{revtex4-2}

\usepackage{url}
\usepackage{txfonts}

\usepackage{amsmath, amssymb}

\usepackage{pstricks}
\usepackage{graphicx}
\usepackage{tikz}
\usepackage[T1]{fontenc}
\usepackage{color}
\usepackage{chemformula}
\usepackage[colorlinks=true,breaklinks=true,allcolors=blue]{hyperref}

\usepackage{tabularx}
\usepackage{multirow}
\usepackage{array}

\usepackage{physics}
\usepackage{xparse}
\usepackage[version=4]{mhchem}
\usepackage[normalem]{ulem}

\newcommand{\phis}{\varphi_\mathrm{S}}
\newcommand{\mf}{m_\mathrm{F}}
\newcommand{\Fperp}{F_{\perp}}
\newcommand{\e}{\mathrm{e}}

\makeatletter
\let\cat@comma@active\@empty
\makeatother

\begin{document}

\title{Observation of sine-Gordon-like solitons in a spinor Bose-Einstein condensate}

\author{Yannick Deller}
\email{sine-Gordon\_Soliton@matterwave.de}
\affiliation{Kirchhoff-Institut f\"{u}r Physik, Universit\"{a}t Heidelberg, Im Neuenheimer Feld 227, 69120 Heidelberg, Germany}%

\author{Alexander Schmutz}
\affiliation{Kirchhoff-Institut f\"{u}r Physik, Universit\"{a}t Heidelberg, Im Neuenheimer Feld 227, 69120 Heidelberg, Germany}%

\author{Raphael Schäfer}
\affiliation{Kirchhoff-Institut f\"{u}r Physik, Universit\"{a}t Heidelberg, Im Neuenheimer Feld 227, 69120 Heidelberg, Germany}%

\author{Alexander Flamm}
\affiliation{Kirchhoff-Institut f\"{u}r Physik, Universit\"{a}t Heidelberg, Im Neuenheimer Feld 227, 69120 Heidelberg, Germany}%

\author{Florian Schmitt}
\affiliation{Kirchhoff-Institut f\"{u}r Physik, Universit\"{a}t Heidelberg, Im Neuenheimer Feld 227, 69120 Heidelberg, Germany}%

\author{\\{Ido Siovitz}}
\affiliation{Kirchhoff-Institut f\"{u}r Physik, Universit\"{a}t Heidelberg, Im Neuenheimer Feld 227, 69120 Heidelberg, Germany}%

\author{Thomas Gasenzer}
\affiliation{Kirchhoff-Institut f\"{u}r Physik, Universit\"{a}t Heidelberg, Im Neuenheimer Feld 227, 69120 Heidelberg, Germany}%

\author{Panayotis G. Kevrekidis}
\affiliation{Department of Mathematics and Statistics, and Department of Physics, University of Massachusetts Amherst, Amherst, 01003, MA, USA}%

\affiliation{Theoretical Sciences Visiting Program, Okinawa Institute of Science and Technology Graduate University, Onna, 904-0495, Japan}

\author{Helmut Strobel}
\affiliation{Kirchhoff-Institut f\"{u}r Physik, Universit\"{a}t Heidelberg, Im Neuenheimer Feld 227, 69120 Heidelberg, Germany}%

\author{Markus K. Oberthaler}
\affiliation{Kirchhoff-Institut f\"{u}r Physik, Universit\"{a}t Heidelberg, Im Neuenheimer Feld 227, 69120 Heidelberg, Germany}%

\date{\today}

\begin{abstract}
We experimentally generate sine-Gordon-like solitons in a spin-1 spinor Bose-Einstein condensate (BEC) utilizing a robust and reproducible local phase-imprinting scheme. We find that the soliton velocity can be tuned by the effective quadratic Zeeman shift. This enables the investigation of controlled soliton interactions, in which we observe the characteristic 
elastic collision behavior of the integrable sine-Gordon model. 
The spatial displacement ---the so-called phase shift--- between incoming and outgoing solitons, the signature of their pairwise interaction,
is found to be in quantitative agreement with numerical spin-1 simulations within the error bars. 
These results establish spinor BECs as a highly controllable experimental platform for studying aspects of the dynamics of sine-Gordon-like models.
\end{abstract}

\maketitle

\textit{Introduction}---Solitons are shape-retaining wave packets that can exist in nonlinear media when the dispersion is compensated by the nonlinear interaction. They are found in a variety of physical systems in nature, e.g., in water, clouds, and even biological systems~\cite{ablowitz2011nonlinear}.
The sine-Gordon (sG) model emerges as a prototypical integrable nonlinear model of soliton dynamics that captures the dynamical behavior of a multitude of physical systems as diverse as Josephson-junction arrays, spin waves in magnetic materials, and lattice dislocations in crystals~\cite{Cuevas-Maraver2014}. 
In addition to its physical relevance, even for systems as simple as coupled torsion pendula~\cite{eilbeck}, and mathematical importance, as on surfaces of constant negative curvature~\cite{mclachlan}, the sG model is one of the cornerstones of the theory of nonlinear waves and solitons.
Examples include kinks and breathers, their elastic wave interactions, and numerous perturbed variants of such solutions~\cite{Kivshar89}.

The sG model has also recently been shown to describe various phenomena in ultracold atomic systems, e.g., in coupled condensates~\cite{Schweigler2017,Bastianello2024sineGordonGHD}, in particular solitons in these systems~\cite{Wybo2023solitons}. 
More concretely, the sG model emerges as the low-energy effective field theory of a weakly interacting ferromagnetic spin-1 Bose-Einstein condensate (BEC) \cite{Siovitz2025} in a specific parameter regime. 
Multi-component BECs have been developed towards a pristine platform for the creation of nonlinear wave patterns, including vector solitons \cite{Bersano2018,Lannig2020}, 
magnetic solitons \cite{Farolfi2020,Chai2020, Rabec2025}, 
as well as Townes solitons \cite{Bakkali_2021}, among
others \cite{Stamper-Kurn2013}.

\begin{figure}
  \centering
    \includegraphics[width=0.4\textwidth]{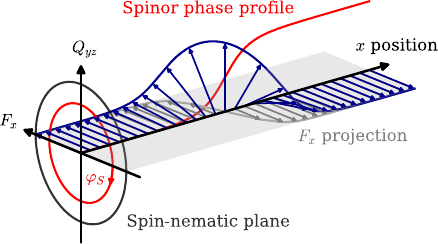}
  \caption{Visualization of the sine-Gordon soliton embedded in the spin-1 system. The soliton is a smooth phase-kink in the spinor phase $\phis$ (see main text for definition), depicted as the red curve. Advancing the spinor phase amounts to a rotation in the spin-nematic plane, which is spanned by the spin observables $F_x$ and $Q_{yz}$. With the phase-kink profile, this leads to a Bloch domain wall structure in the transversal spin. Here we consider the case where the spin is completely polarized along the $F_x$-direction for ease of visualization.}
    \label{fig:figure_1}
\end{figure}

In this letter, we report on the experimental generation of sine-Gordon-like solitons in a spinor BEC by employing a robust and reproducible phase-imprinting scheme. 
In contrast to the aforementioned experimental works, these solitons correspond to magnetic domain walls, as can be seen in~Fig.~\ref{fig:figure_1}, similar to the theoretical proposals introduced in~\cite{Schmied2020, Yu2021, Yu2022, Yu_CoreStructure2022,Yu2024}. \\

\begin{figure}
  \centering
    \includegraphics[width=0.49\textwidth]{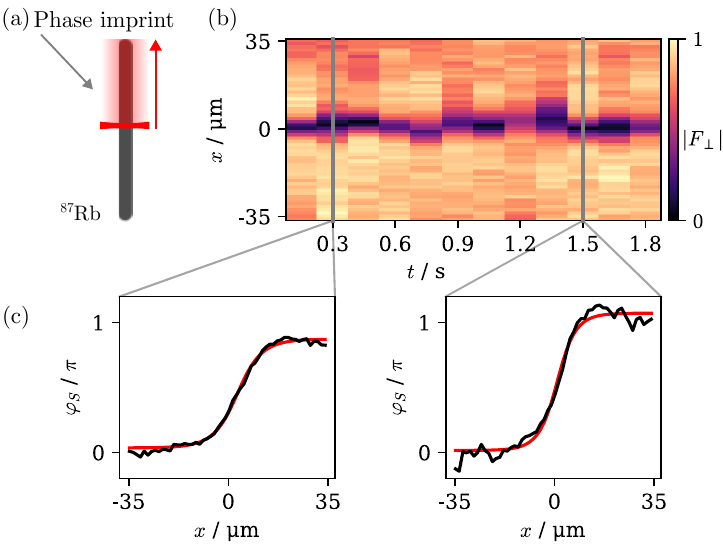}
  \caption{Single sine-Gordon soliton in spinor phase $\phis$ and transversal spin length $|F_{\perp}|$. (a) The one-dimensional system is initially prepared homogeneously in $F_x$. 
  Applying a local magnetic field via the vector Stark shift of a steerable laser beam to one half of the system advances the spinor phase, thereby rotating the spin in the spin-nematic plane spanned by $F_x$ and $Q_{yz}$.
  The smooth kink in the spinor phase profile is created by the Gaussian edge of the imprint laser, see also Fig.~\ref{fig:simulation_spinorphase_slices} for a numerical study. 
  (b) Experimental data of isolated non-moving solitons for up to two seconds evolution time after preparation, showing the length of the transversal spin order parameter $F_{\perp}$. The soliton is visible as a dip in the center of the system that remains static in space. 
  (c) Spinor-phase profile at an early time (left) and at a late time (right) with fitted analytical sine-Gordon kink profile given by Eq.~\eqref{eq:spinorphasekink}. The soliton retains the functional form of the sine-Gordon kink for its full evolution time, as can be seen from the fit to the data (red line).}
    \label{fig:figure2}
\end{figure}

\textit{Experimental system}---We realize a quasi one-dim\-ensional (1D) spin-1 BEC of $\sim 10^5$ \ch{^{87}Rb} atoms in a box trap, with a length of $\sim100 \,\text{µm}$, in a weak magnetic offset field  of $\sim 0.9 \, \mathrm{G}$. 
The spin-1 system manifests itself in the Zeeman states $\mf= 0, \, \pm 1$ of the ground-state manifold $F=1$ . 
The Hamiltonian density of the spinor BEC, in the absence of a trapping potential, is given by \cite{Stamper-Kurn2013}
\begin{equation}\label{eq:hamiltonian}
    H = \frac{\hbar^2 |\partial_x \boldsymbol{\psi}|^2}{2 M} + q\, \boldsymbol{\psi}^{\dagger} f_z^2 \boldsymbol{\psi} + \frac{c_0}{2} |\boldsymbol{\psi}^{\dagger} \boldsymbol{\psi}|^2 + \frac{c_1}{2} |\boldsymbol{\psi}^{\dagger} \boldsymbol{f} \boldsymbol{\psi}|^2\,.
\end{equation}
Here $\boldsymbol{\psi} = (\psi_1, \psi_0, \psi_{-1})^T$, with $\psi_{\mf} = {n_{\mf}^{1/2}} \, e^{\mathrm{i} \varphi_{\mf}}$ where $n_{\mf}$ is the atomic density in the sub-level,
is the mean-field spinor wavefunction, 
$\boldsymbol{f} = (f_x, f_y, f_z)^T$ 
are spin-1 matrices \cite{Kawaguchi2012}, the density-density and spin-spin interaction constants are denoted by $c_0$ and $c_1$, respectively, and $M$ is the mass of rubidium. The quadratic Zeeman shift, which is the energy difference between the $\mf=\pm 1$ levels and the $\mf=0$ level, is denoted by $q$. For all experiments reported here, we prepare the system with an approximately flat atomic density $n=\boldsymbol{\psi}^{\dagger}\boldsymbol{\psi}$, and as the relation $|c_0| \gg |c_1|$ holds for \ch{^{87}Rb}, $n$ stays approximately constant for the entire time evolution.

In the following, we first focus on the case where the quadratic Zeeman shift vanishes. 
Dynamics of the spin degree of freedom may, in general, still be complicated.  
However, one can dramatically reduce complexity by considering normalized states of the form 
\begin{equation}
\boldsymbol{\psi}_{\mathrm{g}} (x,t) 
=  \frac{1}{2}
\begin{pmatrix}  
\e^{- \mathrm{i} \phis(x,t)/2}\\ 
{\sqrt{2}}\, \e^{\mathrm{i} \phis(x,t)/2}\\ 
\e^{- \mathrm{i} \phis(x,t)/2} \end{pmatrix} \,.
\end{equation} 
Here, $\phis$ is the so-called spinor phase, which is defined as $\phis = \varphi_0 - (\varphi_1 + \varphi_{-1}$) /2, where $\varphi_{m_F}$ denotes the phase of the respective Zeeman state. For $\phis=0$, this corresponds to an elongated spin along $F_x$, which is a mean-field ground state for Eq.~\eqref{eq:hamiltonian}, as rubidium features ferromagnetic interactions, i.e. $c_1 < 0$. The low-energy excitations of the system are captured by the spinor phase $\phis$, which follows a sG equation~\cite{Siovitz2025}, see also \footnote{One can transform Eq.~\eqref{eq:sG_equation} into a sine-Gordon equation in standard form by using the transformation $\bar{x} = x \, \sqrt{2 \lambda / c_s^2}$, $\bar{t} = t \, \sqrt{2 \lambda}$ and $\bar{\varphi}_s = 2 \varphi_s$.},
\begin{equation}\label{eq:sG_equation}
    (\partial_t^2 - c_\mathrm{s}(\bar{q})^2 \, \partial_x^2 ) \, \phis 
    + \lambda \, \mathrm{sin} (2 \phis) = 0.
\end{equation}
The propagation speed $c_\mathrm{s}$ of the free sG model and the coupling $\lambda$ depend on the parameters of the spin-1 Hamiltonian via $c_\mathrm{s}(\bar{q}) = \sqrt{2(1-\bar{q}^2)}$ and $\lambda = 2(1-2\bar{q}^2)$, where $\bar{q} = q / 2 n |c_1|$. 
Here, space and time are given in units of the spin healing length $\xi_\mathrm{s} = \hbar/\sqrt{2 M n |c_1|}$ and spin healing time $t_\mathrm{s} = \hbar / n |c_1|$. 
The value $c_\mathrm{s}(\bar{q}=0)=\sqrt{2}$ is equal to the spin speed of sound (in SI units $c_{\mathrm{s}, \mathrm{SI}} = \sqrt{2} \, \xi_s/t_\mathrm{s}  =\sqrt{n |c_1|/M} = \hbar / 2 M \bar{\ell} \sim 70\,\text{µm}/s$ for our experimental parameters), and the width of the soliton is given by $\bar{\ell} = \xi_s/\sqrt{2}$. \\
For $q>0$, the equation of motion for $\phis$ is given by the double-sG equation, which contains a term proportional to $\mathrm{sin}(4 \,\phis)$. However, for the experiments reported below, the double-sG term is negligible even for $q \neq 0$, when using the experimental parameters and the coefficients derived in \cite{Siovitz2025}.
Considering states of this form reduces the complexity of the three coupled complex-valued  differential equations arising from Eq.~\eqref{eq:hamiltonian} to a single differential equation of one real scalar field \cite{Rutkevich2024}, which is known to have 1D solitonic solutions.

In order to deterministically generate solitons, we first prepare our system homogeneously in a fully elongated transverse spin state along $F_x$. 
We use a combination of global state transfer to $F=2$ and a local effective magnetic field \cite{Lin2009} induced with a steerable laser beam at the tune-out wavelength \cite{Schmidt_2016} to imprint the desired spinor phase profile given by the following Eq.~\eqref{eq:spinorphasekink}
\begin{align}
    \phis(x,t) = A \hspace{0.1cm} \textrm{arctan} \left[\mathrm{e}^{(x-vt-x_0)/ \ell}\right] +C .
    \label{eq:spinorphasekink}
\end{align}
with minimal disturbance of the total density, cf.~Fig.~\ref{fig:figure2}~(a). Here, $A$ is the amplitude, $\ell$ the width, $x_0$ the initial position, $v$ the velocity, and $C$ the offset-phase of the soliton.
This direct phase imprinting is similar to the one employed in \cite{Farolfi2020, Chai2020} to generate magnetic solitons, however, we directly address a different phase degree of freedom, i.e., $\phis$, by holding the atoms in a suitable configuration of hyperfine states, see Fig~\ref{fig:Imprint_multibeam}.

\textit{Dynamics of a single soliton}---As shown in Fig.~\ref{fig:figure_1}, the imprint of the sG soliton kink-profile in the spinor phase, which interpolates in a localized fashion between $0$ and $\pi$, rotates the $F_x$ spin into $-F_x$, thereby generating a localized maximum in the quadrupole degree of freedom $Q_{yz}$. The latter is a rank-$2$ operator, which is needed in addition to the rank-$1$ spin operators introduced above to describe the spin-1 system \cite{Hamley2012}, and is defined as $Q_{yz} = \boldsymbol{\psi}^{\dagger} (f_y f_z + f_z f_y)\boldsymbol{\psi} /2$. This results in a localized reduction of $F_x$. Experimentally, we measure $F_x$ and $F_y$ simultaneously using a POVM-readout scheme \cite{Kunkel2019} and extract $F_{\perp} = F_x + \mathrm{i} F_y$, as the spin orientation in the transverse plane is randomized globally in each experimental realization as a result of spin precession due to the linear Zeeman effect in the fluctuating magnetic offset field. As $|\Fperp|$ is the magnitude of the transversal spin length, a localized reduction and subsequent sign flip of $F_x$ leads to a localized dip in $|\Fperp|$, as can be seen in Fig.~\ref{fig:figure2}~(b). In addition, we are able to simultaneously measure $F_x$ and $Q_{yz}$, which allow us to determine the spinor phase from $\tan\phis=F_{x}/Q_{yz}$. 

\begin{figure}
    \centering
    \includegraphics[width=0.49\textwidth]{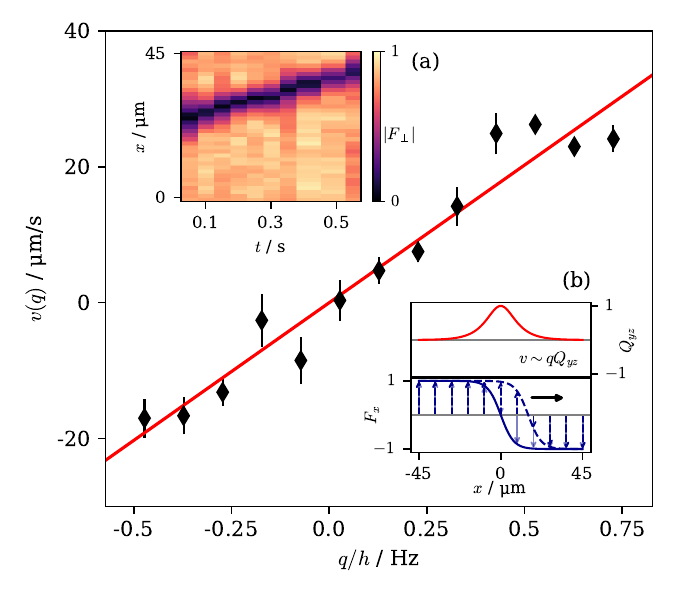}
    \caption{Linear dependence of the soliton velocity $v$ on the quadratic Zeeman shift $q$. In the inset (a), the experimental data for (post shift) $q/h=0.53 \,\mathrm{Hz}$ are shown. In inset (b) a conceptual visualization of the moving soliton in the framework of the spin-continuity equation is shown, where a $q$-dependent source term co-moves with the soliton.}
    \label{fig:figure3}
\end{figure}
The dynamics in the transverse-spin length after imprinting a phase profile close to an isolated sG soliton is shown in Fig.~\ref{fig:figure2}~(b). We tune the quadratic Zeeman shift $q$ by an off-resonant microwave dressing, so that the soliton does not move. 
The experimental viability of reliably varying $q$ enables control over the velocity of the soliton, as will be discussed in the following in detail. 
Fig.~\ref{fig:figure2}~(c) furthermore shows that the soliton retains the analytical functional form in the spinor phase during the entire time evolution, which is given by Eq.~\eqref{eq:spinorphasekink}.\\
However, both the width $\ell$ and the amplitude $A$ of the kink undergo oscillations, as we do not perfectly imprint the static length scale (to a good approximation given by the spin-healing length $\xi_s$, which is $\sim 7\, \text{µm}$ for the experiments reported here, divided by $\sqrt{2}$) and the static amplitude $A=2$ of the soliton. The resolution of the imprint is $\sim 4 \text{µm}$.
This dynamical behavior due to imperfect preparation is confirmed in mean-field simulations, see Fig.~\ref{fig:numericalstudy_widthosci} for more information. In this way, we corroborate, that the velocity of the soliton is independent of its initial width, even as the width undergoes oscillations. The experimental imprinting technique is therefore robust enough for a systematic study of the soliton's velocity.\\

Varying the quadratic Zeeman shift $q$ changes the soliton's velocity, as discussed in the following. 
We record the time evolution of the observable $F_{\perp}$, for each $q$, for up to $1\,$s and extract the velocity $v$ from a linear fit to the soliton's position, cf.~Fig.~\ref{fig:figure3} (inset a) for an example data set. 
We find that the velocity varies linearly in $q$, $v(q) = \alpha\,q/h$, with $\alpha = (40\pm 3) \, \text{µm}$/(s$\cdot$Hz), which we extract from a linear fit, see Fig.~\ref{fig:figure3}.
We subtract an offset from our experimental $q$ value, such that $v(q=0) = 0$, which is motivated further below. The value for $q$, at which the soliton is non-moving also coincides with minimal oscillations of the background transversal spin length and is therefore corroborated by an independent experimental signal.
Remarkably, the soliton remains stable for up to $15\,$s even at non-zero velocities and despite multiple reflections at the system's boundary during its lifetime.

The linear dependence of the velocity on $q$ can be understood both in the full spin-1 model and within the effective sG theory. 
The width $\ell$ of a sine-Gordon kink is in general a free parameter and determines the velocity via $c_\mathrm{s}/\ell = \sqrt{2\lambda} \, \gamma$, with Lorentz-factor $\gamma = (1-v^2/c_\mathrm{s}^2)^{-1/2}$ and coupling $\lambda$. 
However, in both experiment and numerical simulations we observe that the width of the soliton in the spin-1 system is, to a good approximation,
given by $\ell =  \xi_\mathrm{s}/\sqrt{2}$, in line with previous numerical studies~\cite{Yu2021}. 
We are therefore able to access only a subclass of kink-excitations of the set of all sine-Gordon solutions, with their width fixed by the density $n$ and coupling $c_1$.  
As we show in
App.~\ref{app:Velocity}, the general sG result is consistent with this finding and both the full spin-1 and the effective sine-Gordon theory yield the relation
\begin{equation}
    {v}/{\ell} = \pm \, {q}/\hbar ,
    \label{eq:velocity}
\end{equation}
between velocity, width, and quadratic Zeeman shift, where the sign differs for kink and anti-kink solutions. 
We find, from the mean width of all experimental data points (we measure $q/h$ in $\mathrm{Hz}$), that $(2 \pi \ell)\approx 32\,\mu$m/(s$\cdot$Hz), which is on the same order of magnitude as the experimentally extracted value for $\alpha$. 
In the spin-1 model, Eq.~\eqref{eq:velocity} results from a spin-continuity equation, where the peak in the quadrupole operator $Q_{yz}$ at the soliton's position acts as a co-moving source that causes the gradual decrease (or increase for an anti-kink) of $F_x$, see inset (b) of Fig.~\ref{fig:figure3} and App.~\ref{app:Velocity}.
It is furthermore worth noting, that the soliton is dynamically stable even for negative $q$, not anticipated from the mean-field phase diagram \cite{Kawaguchi2012}, where, for $q<0$, excitations in $F_z$ are expected. 
We presume, however, that the soliton remains metastable, as global conservation of $F_z$ prevents it from decaying.

It is, moreover, possible to adjust the microwave dressing for different densities such that the soliton does not move, see App.~\ref{app:densityDependentShift} for more information.
However, the value of $\alpha / \overline{\ell}$ remains, within errors, the same for all measured atomic densities, where $\overline{\ell}$ is the mean width of the soliton averaged over all realizations, which in turn also depends on the density through the spin-healing length.

\textit{Kink-anti-kink ($K$-$\overline{K}$) collisions}---The experimental control over the velocity enables us to study $K$-$\overline{K}$ collisions. For this, we imprint a rectangular phase profile with Gaussian edges at the center of the condensate, which generates a $K$-$\overline{K}$-pair at a distance of $\sim35\, \text{µm}$. 
\begin{figure}
    \includegraphics[width=0.49\textwidth]{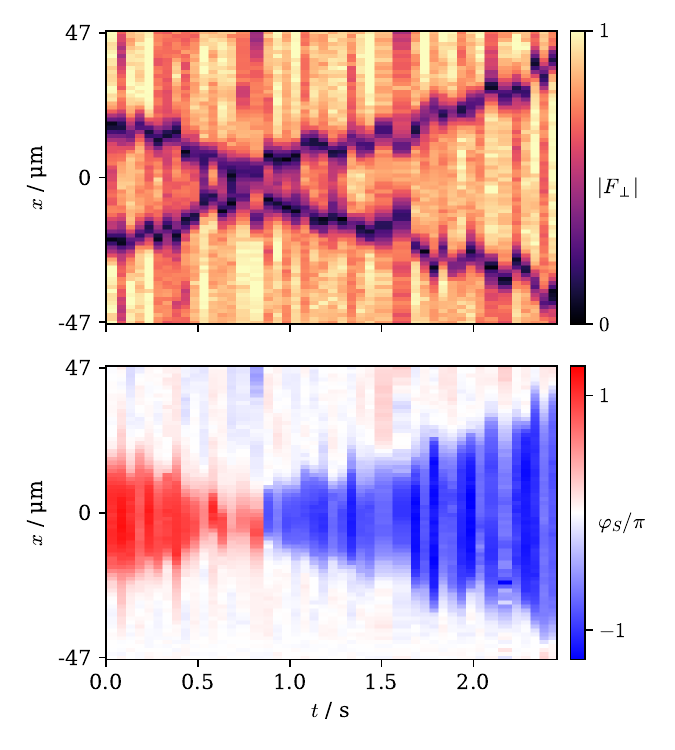}
    \caption{Kink-anti-kink collision measured simultaneously in different observables. In the upper panel, the collision is shown in the transverse spin length $|F_{\perp}|$. In the lower one, a gradual $2\pi$ phase change is visible in the spinor phase during the collision, which is in accordance with predictions for the sine-Gordon model. The quadratic Zeeman shift is given by $q/h \sim 0.5\,$Hz for this experiment.}
    \label{fig:figure4}
\end{figure}
As can be seen in Fig.~\ref{fig:figure4}, we observe a gradual 2$\pi$ change in the spinor phase during the collision. 
This is in accordance with, but not limited to, the analytical prediction for the sine-Gordon model \cite{Cuevas-Maraver2014}, as such a gradual phase change can be observed even in linear wave equations.
We simultaneously measure the transversal spin profile $F_{\perp}$ and the spinor phase $\phis$, see App.~\ref{app:POVMReadout} for details.

\begin{figure}
    \centering
    \includegraphics[width=0.49\textwidth]{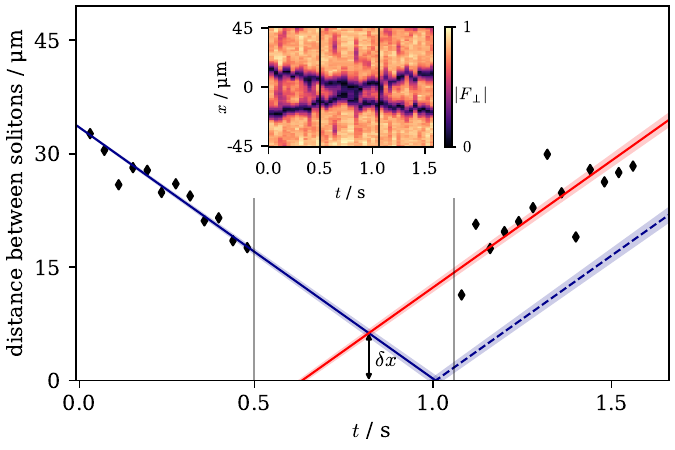}
    \caption{
    Phase-shift extraction in the elastic regime. The obtained velocity for the incoming solitons is $v_\mathrm{in}=\left(16.8\pm0.3\right)\, \text{µm}/\mathrm{s}$. In order to extracted the phase shift we assume energy conservation which yields $\delta x = (6.3 \pm 0.6)\, \text{µm}$. This is consistent within the error with the value extracted from spin-1 numerical 1d simulations, which is $6.2\, \text{µm}$.
    }
    \label{fig:figure5}
\end{figure}

Elastically colliding solitons experience a so-called phase shift due to interactions during the collision, that is, a positional shift that the soliton experiences due to the interaction during the collision, so that its trajectory deviates from the one expected for a hard core collision.
In order to extract this phase shift, we determine the relative distance of the solitons from a fit to the data of the transverse spin length, see Fig.~\ref{fig:figure5}, where we choose $q/h \sim 0.3 \,\mathrm{Hz}$.
We fit a linear function to the relative distances both before and after the collision region, which we define as the time interval in which the soliton profiles are separated by less than three times their width, and where their individual positions are not well defined any more. We find that the collision is highly elastic with $v_\mathrm{out}/v_\mathrm{in} \approx 0.9$, such that it is feasible to meaningfully extract phase shifts. By assuming perfect elasticity, i.e. by using the same slope for the relative distances of the outgoing solitons as for the incoming ones, we find that the phase shift of the collision data shown in Fig.~\ref{fig:figure5} is $\delta x = (6.3 \pm 0.6)\, \text{µm}$.

Furthermore, we extract the phase shift from one-dimensional spin-1 simulations, which yields $\delta x_{\mathrm{num}}= 6.2 \, \text{µm}$ for the same incoming velocity. 
Both values agree quantitatively and are positive, reflecting the attractive interaction in the sG model.
This observation is on the same order of magnitude as the phase shift obtained from the analytical prediction of the sG model~\cite{eilbeck, Kivshar89} by employing the mapping between the spin-1 model and the sG effective field theory from~\cite{Siovitz2025}, see App.~\ref{app:PhaseShift}. 
However, the value predicted by the effective model is $\delta x_{\mathrm{sG}} = 14.8\, \text{µm}$, and thus more than twice as large as the observed phase shift.

We attribute this deviation to the fact that the collision event violates the assumptions made in deriving the effective sG model \cite{Siovitz2025}. 
Nevertheless, it is remarkable that our observations can be captured by choosing a sG coupling $\sqrt{\lambda}=2.5$, which gives excellent agreement between the sG and the numerical spin-1 predictions, as shown in Fig.~\ref{fig:numerical_phase_shift} in App.~\ref{app:Numerics} for the entire range of monitored kink speeds. 
These observations suggest that the phase shift can be accounted for by adjusting the coupling constant, which is left for further theoretical investigation.

\textit{Discussion and outlook}---We have deterministically generated sine-Gordon-like solitons in a 1D spin-1 Bose-Einstein condensate and studied their dynamics, finding that the soliton velocity can be controlled by adjusting the effective quadratic Zeeman shift $q$. 
Furthermore, we have generated soliton pairs and investigated their behavior upon collision. 
We were able to extract a spatial phase shift, demonstrating that the experiment's stability and control are sufficient for studying the interaction between the solitons. 
The experimentally extracted value of the phase shift is in accordance with our numerical spin-1 simulations.

In turn, this study paves the way for numerous extensions of imminent interest, given in particular the versatility of the experimental system. 
One natural question associated with our control of the wave speed concerns its potential limitations analogous to the relativistic limit.
The sG equation contains both second-order spatial and time derivatives, reflecting the non-dispersive sound excitations of the underlying non-linear Schr\"{o}dinger equation, with a linear relation $\omega\sim c_\mathrm{s} k$ between energy and momentum (for large values of $k$). 
It is natural to inquire about the extension of our $q$-$v$-diagram near the limit $v\to c_\mathrm{s}$, and the associated relevance of the Lorentz factor.\\
Moreover, sG is unique among classical (homogeneous) 
{\it continuum} field theories~\cite{BMW94} since it has breather solutions.
Generating and examining breathers and their interactions~\cite{hiroshima}
is of particular interest in its own right.
Going beyond the sG realm, it turns out that the spin system provides
a path towards non-integrable generalizations 
of the sG description. Indeed, the spin-1 GP model
is related to the non-integrable {\it double} sG model~\cite{Siovitz2025} 
that has been the subject of intense theoretical scrutiny in the nonlinear-waves community~\cite{Campbell1986}. This is due to its numerous
intriguing features including kink internal modes and their
intrinsic breathing, fractal (so-called $n$-bounce)
kink-anti-kink collisions; see, e.g.,~\cite{Bai2025} for a recent numerical investigation of similar phenomena in spinor BECs. 
The present platform opens many possibilities for studying such field theories in future theoretical, numerical and
experimental studies.\\

The authors thank Marius Sparn, Hannes K{\"o}per and Anna-Maria Gl{\"u}ck for discussions and Felix Klein for experimental assistance during the beginning of the project.
They acknowledge support 
by the Deutsche Forschungsgemeinschaft (DFG, German Research Foundation), through 
SFB 1225 ISOQUANT (Project-ID 273811115), 
grant GA677/10-1, 
and under Germany's Excellence Strategy -- EXC 2181/1 -- 390900948 (the Heidelberg STRUCTURES Excellence Cluster), 
and by the state of Baden-W{\"u}rttemberg through bwHPC and the DFG through 
grants 
INST 35/1503-1 FUGG, INST 35/1597-1 FUGG, and INST 40/575-1 FUGG
(SDS, Helix, and JUSTUS2 clusters).
This material is partially based upon work supported by the U.S. National Science Foundation under the award PHY-2408988 (PGK). This research was partly conducted while P.G.K. was 
visiting the Okinawa Institute of Science and
Technology (OIST) through the Theoretical Sciences Visiting Program (TSVP). 
This work was also supported by the Simons Foundation
[SFI-MPS-SFM-00011048, P.G.K.].



\begin{appendix}

\onecolumngrid
\vspace{1cm}
\begin{center}
\textbf{APPENDIX}
\end{center}
\setcounter{equation}{0}
\setcounter{table}{0}
\makeatletter

In this Appendix, we provide further details on the experimental system, numerical simulations of the spin-1 system and analytical calculations in both spin-1 and sG theory to support the results in the main text.

\section{Experimental system}\label{app:detailsExperiment}
We prepare a quasi-one-dimensional BEC of $\sim 10^{5}$ $\ch{^{87}}$Rb atoms in a box-like nearly homogeneous 
trapping potential. The trap is realized by using a red-detuned laser beam with a wavelength of $1030\,\mathrm{nm}$, resulting in a quasi-1D harmonic trap ($w_r = 2 \pi \times 170\,\mathrm{Hz}$ and $w_l = 2 \pi \times1.6\,\mathrm{Hz}$). The box walls are realized by two blue-detuned steerable laser beams with a wavelength of $760\,\mathrm{nm}$, see~\cite{Lannig2020} for details.
All experiments are performed in an actively stabilized homogeneous magnetic offset field of $B =0.894\,\mathrm{G}$, which gives rise to the quadratic Zeeman shift of $q(B)/h \sim 58\,\mathrm{Hz}$. By applying off-resonant microwave dressing, we couple, in leading order, the $\ket{F=1,\mf=0}$ and $\ket{F=2,\mf=0}$ levels. This allows us to effectively tune $q$ into a regime on the order of the spin-spin interaction, $q \sim n |c_1|$. At the start of every experimental sequence reported here, we prepare the BEC close to the easy-plane ground state of $q=0$, which corresponds to an elongated transversal spin  with $|\Fperp| \approx 1$. For this, all atoms are first transferred to $\ket{F=1,\mf=0}$, followed by a $\pi/2$-radio frequency (RF) pulse which rotates the state around the $F_x$-axis. In order to align the spin on the $F_x$-axis, we adjust the global spinor phase to $\phis =(0 \mod 2\pi)$, which is defined as $\phis = (\varphi_1 + \varphi_{-1})/2 - \varphi_0$. This is implemented by two subsequent MW $\pi$-pulses between $\ket{F=1,\mf=0}$ and $\ket{F=2,\mf=0}$ with a relative phase difference between the first and second pulse.

\section{Experimental generation of solitons}\label{app:detailsimprint}

\begin{figure*}
    \centering
    \includegraphics[width=\linewidth]{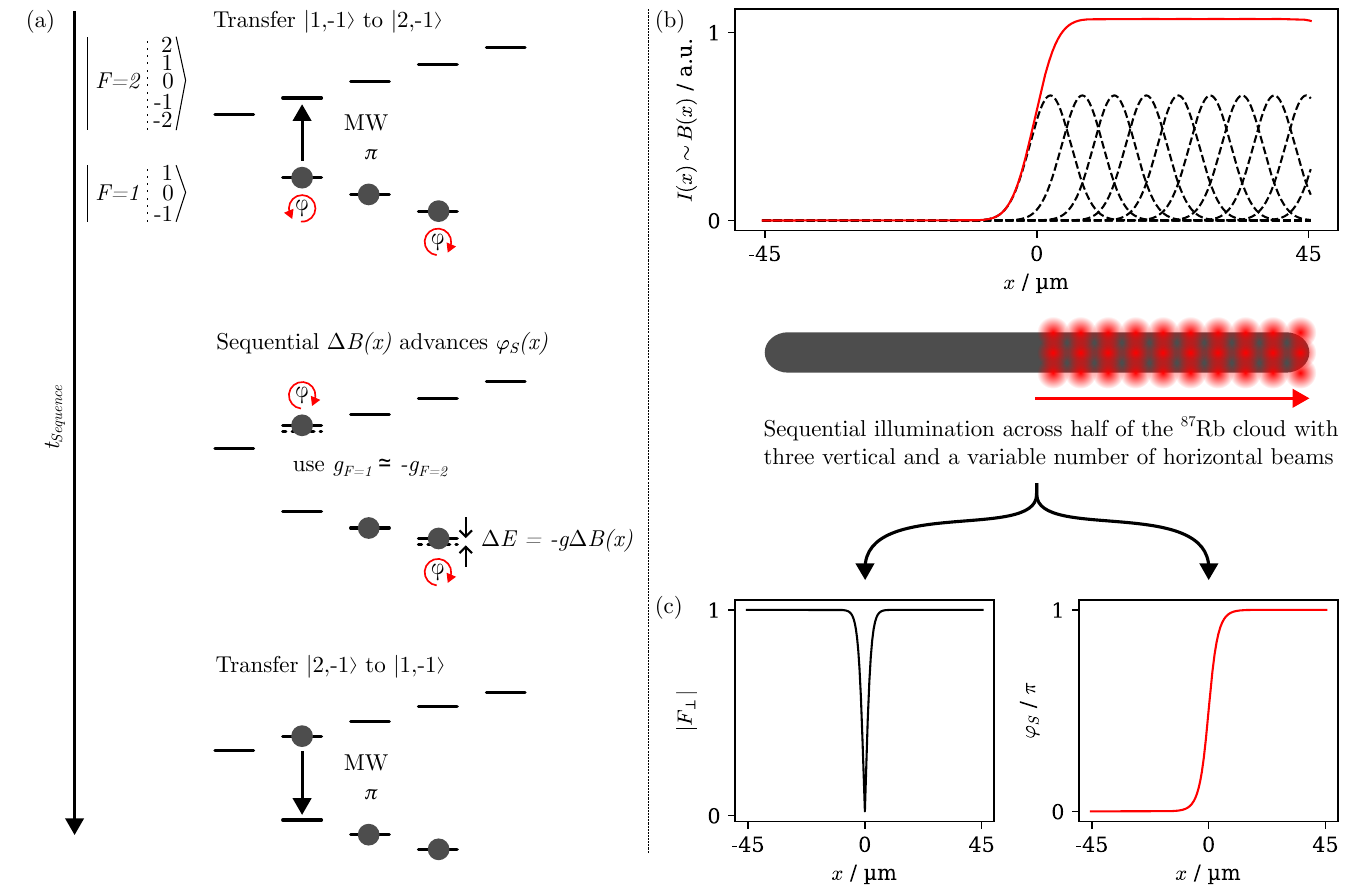}
    \caption{Scheme for the experimental preparation of a sine-Gordon soliton: (a) To advance the spinor phase as the phase difference between $\ket{F=1,\mf=0}$ and $\ket{F=1,\mf=\pm 1}$, we use the opposite sign of the g-factor in $F=1$ and $F=2$. For this, the complete population in $\ket{F=1,\mf=-1}$ is transferred to $\ket{F=2,\mf=-1}$ with a microwave pulse. Holding the atoms in this configuration allows us to locally apply an additional effective B-field by using the vector Stark shift, which advances the spinor phase proportional to the intensity of the laser beam. This is done sequentially over one half of the \ch{^{87}Rb} cloud using two crossed AOD's, as shown in (b). Then, the population in $\ket{F=2,\mf=-1}$ is transferred back to $\ket{F=1,\mf=-1}$. (b) Since the vector Stark shift is proportional to the intensity of the laser beam, the single beams can be added up close to a sine-Gordon soliton profile, (c) resulting in a $\pi$ kink in the spinor phase and a dip in the transversal spin length $|\Fperp|$.}
    \label{fig:Imprint_multibeam}
\end{figure*}

In order to prepare a soliton in the spinor phase, we have to locally manipulate the phase difference between the \(m_F = 0\) and $m_F = \pm 1$ components. This is realized via sequential illumination of the BEC with a steerable focused laser beam at the tune-out wavelength of $790.032 \,\mathrm{nm}$. By choosing this wavelength, the laser induces an additional local magnetic field $\Delta B(x)$ via the vector Stark shift, without causing density excitations~\cite{Pruefer2022}. The additional magnetic field locally manipulates the phase evolution of the $m_F = \pm1$ components, which would lead to a locally advancing Larmor phase $\varphi_L =  \varphi_1 - \varphi_{-1}$. In order to utilize this concept for local spinor phase rotations, we make use of the relation $g_{F=1} \approx - g_{F=2}$, which is a property of the rubidium electronic ground state (but also holds for other alkali species).
Following the sequence shown in Fig.~\ref{fig:Imprint_multibeam}~(a), 
all atoms in the $\ket{F=1, \mf=-1}$ component are transferred globally to $\ket{F=2, \mf=-1}$ by a microwave pulse and held there for a fixed duration, during which the BEC is locally illuminated. Consequently, the condensate phases of $\ket{F=2, \mf=-1}$ and $\ket{F=1, \mf=1}$ are advanced in the same direction. As $\varphi_0$ is unaltered, this effectively advances the spinor phase in a magnetic bias field $B(x)$. During the hold time $\tau$, the spinor phase $\phis$ is locally advanced by $\Delta \phis \sim \Delta B(x) \tau$. We use a setup of two crossed acousto-optic deflectors (AODs) to steer the imprint laser beam over the atom cloud. In order to avoid interference, we turn off the driving signal of the AODs between individual pulses, such that only one radio-frequency tone per pulse is present in each AOD crystal. In this way we apply grids of up to $12$ horizontal and $3$ vertical sequential laser pulses. Consequently, the spatial profile of the spinor phase is directly proportional to the sequentially applied intensity profile (Fig.~\ref{fig:Imprint_multibeam}~(b)), while the global offset value of the spinor phase is adjusted by the hold time. After the local phase imprint we transfer all atoms from $\ket{F=2,\mf=-1}$ back to $\ket{F=1, \mf=-1}$.\\
This method allows us to imprint any functional form in the spinor phase which can be approximated by a sum of Gaussians. While illuminating one half of the condensate sequentially with equal amplitudes, it results in a sine-Gordon soliton in the spinor phase $\phis$ and a dip in the transverse spin length $|\Fperp|$, shown in Fig.~\ref{fig:Imprint_multibeam}~(c). The phase kink stems from the Gaussian edge of the outermost imprint beam, which has a waist of $\sim 7.6 \,\mu \mathrm{m}$. We optimize the longitudinal distance of the sequentially applied Gaussian beams such that the resulting intensity is as homogeneous as possible on the illuminated half of the atomic cloud. 

\begin{figure}[]
    \centering
    \includegraphics[width=0.5\linewidth]{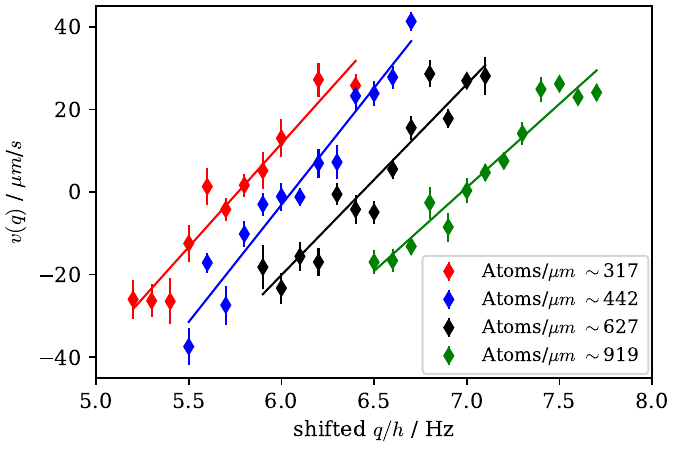}
    \caption{Velocity data for different atomic densities. When varying the atomic density (different colors), one has to apply a different microwave dressing to realize a non-moving soliton. We have fitted a linear function to the data for each atomic density.}
    \label{fig:velocities_diffDensities}
\end{figure}

\begin{figure}[]
    \centering
    \includegraphics[width=0.5\linewidth]{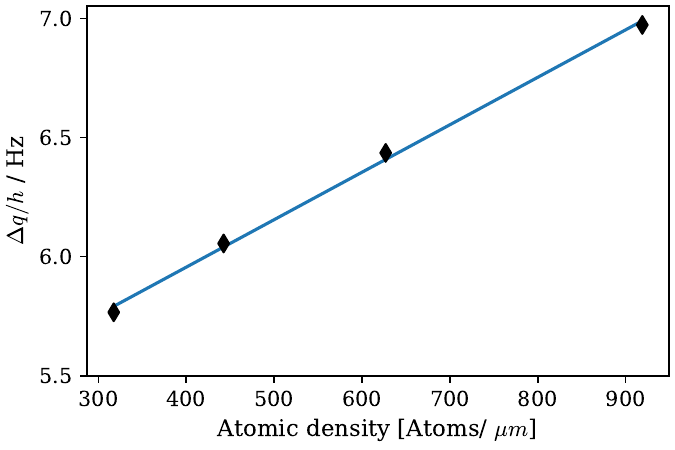}
    \caption{The $x$-intercept from linear fits to the data presented in Fig.~\ref{fig:velocities_diffDensities}, which gives the density dependent $q$-shift $\Delta q$. The blue line is a linear fit to the data.}
    \label{fig:qshift_diffDensities}
\end{figure}

\section{Density dependent q-shift}\label{app:densityDependentShift}
In the following, we give further details about the density dependence of the $q$-shift, which was observed for moving solitons. 
Fig.~\ref{fig:velocities_diffDensities} shows velocity data for different atomic densities. The velocities are not rescaled with the mean soliton width, therefore the slopes do not coincide.
We extract the density-dependent offset $\Delta q$ from the linear fits, and the results are shown in Fig.~\ref{fig:qshift_diffDensities} with a linear fit to the data.
The density-dependent offset is not captured in one-dimensional spin-1 mean-field simulations and is also not present in the analytical calculations based on the spin continuity equation. We have also performed numerical simulations in a three-dimensional trapping configuration close to the experimental one to exclude that the shift is due to the finite transversal extent of the condensate. A possible source for the shift could be dipolar interactions or beyond-mean-field effects, that were not included so far in the numerical simulations.

\section{Width oscillation of the soliton}\label{app:WidthOscillations}
\begin{figure}
  \centering
    \includegraphics[width=0.5\textwidth]{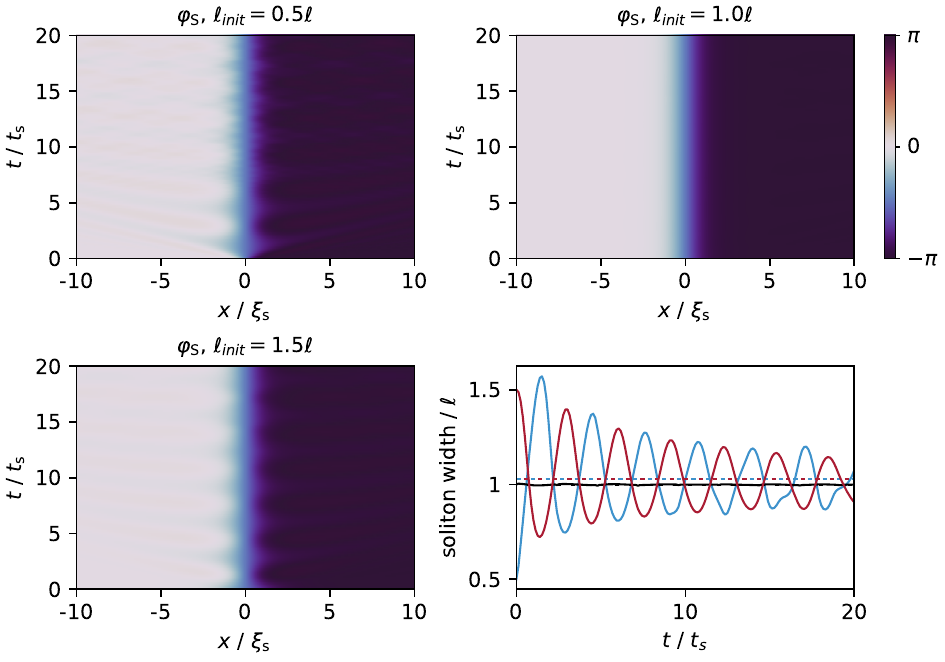}
  \caption{Numerical simulation of soliton dynamics for different initial widths in order to confirm robustness of experimental imprint to imperfections, taken from~\cite{schaefer2025}.}
    \label{fig:numericalstudy_widthosci}
\end{figure}

\begin{figure}
  \centering
    \includegraphics[width=0.5\textwidth]{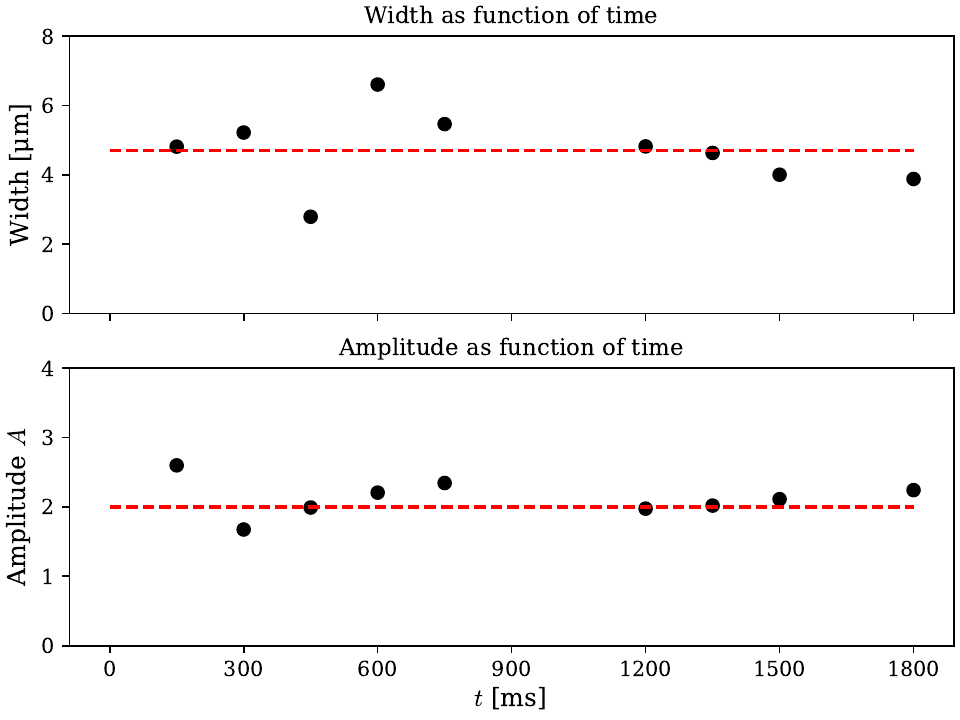}
  \caption{Widths and amplitudes of the soliton ---as expressed in Eq.~(4)
  of the main text--- for the data reported in Fig. 2 of the main text for the full time series The dashed lines correspond to the mean width and the analytically predicted value for $A$, respectively.
  }
    \label{fig:figure_RL11}
\end{figure}

The soliton has a static width, as discussed in \cite{Yu2021}, and confirmed by numerical simulations shown in Fig.~\ref{fig:numericalstudy_widthosci}. When deviating from its static value in the initial condition, which is given by $\ell = \xi_s/\sqrt{2}$, the width shows damped oscillations, which, however, leave the soliton velocity unchanged. Fig.~\ref{fig:figure_RL11} shows the time evolution of the width $\ell$ and amplitude $A$ corresponding to the data shown in the main text in Fig.~\ref{fig:figure2}.

\section{POVM readout for Dual Phase Extraction}\label{app:POVMReadout}
\label{subsec:DuPR}
\begin{figure*}
    \centering
    \includegraphics[width=\linewidth]{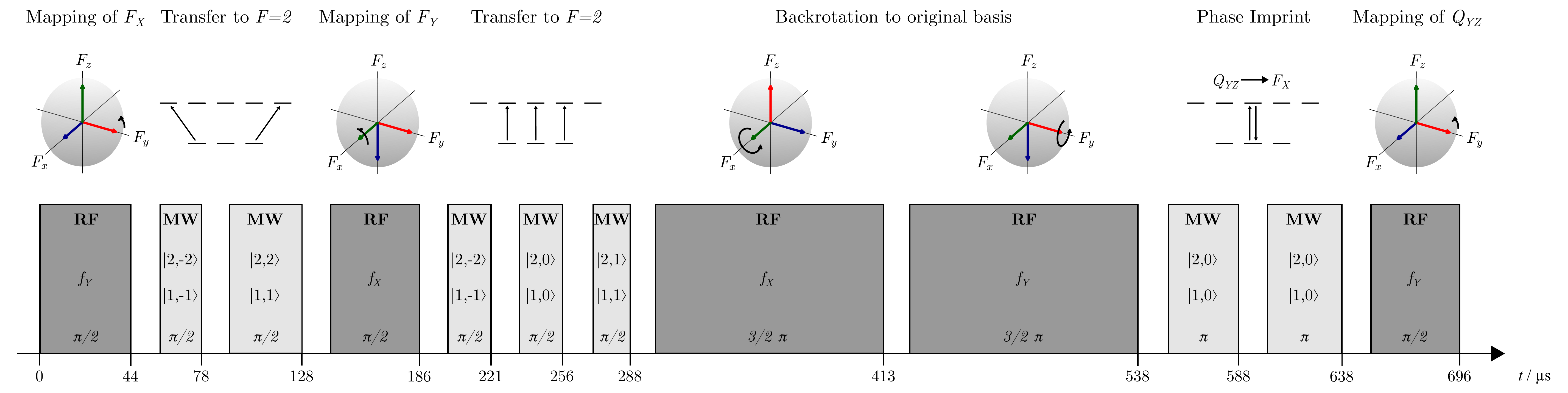}
    \caption{Dual-Phase readout scheme~\cite{2025FlammNovel}: Following the readout method first presented in~\cite{Kunkel2019}, the state is first rotated around the $f_y$ axis, to further transfer $\nicefrac{1}{3}$ of the population from $\ket{F=1,\mf=\pm 1}$ in the F=2 manifold $\ket{F=2,\mf=\pm 2}$. This will store the $F_x$ observable in $F=2$, and it is unaltered as we apply hyperfine-selective RF rotations. Then, after a rotation around $f_x$ in the $F=1$ components, the $F_y$ observable is stored the same way in $\ket{F=2,\mf=\pm 1}$. In addition to that, $Q_0$ can later be extracted via the difference in population of $\ket{F=1,\mf=0}$ and $\ket{F=2,\mf=0}$. Two RF pulses rotate the state to the starting orientation, before two $\pi$-pulses map $Q_{yz}$ to $F_x$. This can now be extracted in the $F=1$ manifold after another $f_y$-rotation.}
    \label{fig:DuPR_Scheme}
\end{figure*}
In the following, we give further information on the readout technique used for taking the data presented in Fig.~\ref{fig:figure4} in the main text. So far, it was possible to extract the transverse spin length $|\Fperp|$ and the spinor phase $\phis$ in two separate measurements, based on previous work~\cite{Kunkel2019}. Expanding on this, we have implemented a new readout scheme that allows us to extract up to four non-commuting observables $ \{ F_x,F_y,Q_{yz},Q_0\} $ in a single measurement~\cite{2025FlammNovel}.\\
For this purpose, we perform global rotations, realized via hyperfine-selective RF pulses in the $F=1$ manifold, and microwave pulses between two magnetic sub-levels $\ket{F=1,m_{1}}$ and $\ket{F=2,m_{2}}$. The population of the respective levels $N_{F,\mf} = a^{\dagger}_{F,\mf} a_{F,\mf}$ change under these rotations, $N^{\prime}_{F,m_F} = U^{\dagger} N_{F,m_F} U $, where $U$ is the unitary rotation operator including all RF- and MW-pulses.
The complete pulse sequence is visualized in Fig.~\ref{fig:DuPR_Scheme}.
We extract the observables from the atomic densities by using the relations
\begin{align}
    F_x &= \frac{1}{N \sin^2{\nicefrac{\phi}{2}}} \left( N_{2-2}^{\prime} - N_{22}^{\prime} \right) \;, \\
    F_y &= \frac{2}{N \cos{\nicefrac{\phi}{2}}} \left( N_{21}^{\prime} - N_{2-1}^{\prime} \right) \;, \\
    Q_{yz} &= \frac{2}{N \cos^2{\nicefrac{\phi}{2}}} \left( N_{11}^{\prime} - N_{1-1}^{\prime} \right)  \;, \\
    Q_0 &= 1 - \frac{4}{N} \left( \frac{N_{20}^{\prime}} {\cos^2{}\nicefrac{\phi}{2}} + N_{10}^{\prime} \right) \;,
\end{align}
which are obtained by direct calculation~\cite{Kunkel2019,2025FlammNovel}. Here, $N$ is the total atom number and $\phi$ is the duration of the first MW pulses for the readout of $F_x$, and is chosen such that $1/3$ of the initial population is transferred to $F=2$ in the measurements reported in the main text.

\section{Numerical spin-1 simulations of soliton dynamics}\label{app:Numerics}
\begin{figure}[]
    \centering
    \includegraphics[width=0.5\textwidth]{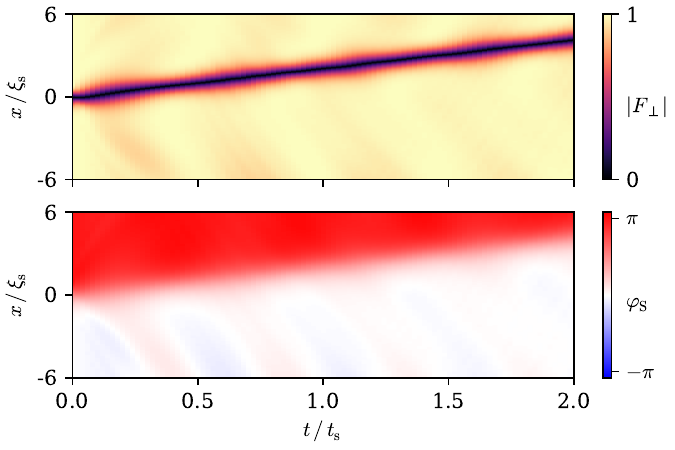}
    \caption{Kink as seen when using the imprint as defined in Eq.~\eqref{eq:kink_simulation} as initial condition for $q=0.3n|c_1|$. We have used the easy-plane ground state at $q=0$ as the initial state, which leads to small background oscillations. Furthermore, one can see width oscillations of the soliton, which are due to the experimental imprint not perfectly matching the static width and functional form of the sG kink. }
    \label{fig:simulation_kink_q03}
\end{figure}

We are numerically simulating the spin-1 Gross-Pitaevskii equation (GPE), which in one spatial dimension is given by
\begin{align}
    \mathrm{i} \hbar \partial_t \boldsymbol{\psi}(x,t) &= \left[-\frac{\hbar^2}{2M}\partial_x^2+V(x)+qf_z^2 \right. \nonumber \\
    &+\left.\frac{c_0}{2}|\boldsymbol{\psi}(x,t)|^2+\frac{c_1}{2}\boldsymbol{F}(x,t)\boldsymbol{f} \right]\boldsymbol{\psi}(x,t), 
\end{align}
where $\boldsymbol{f}$ is a vector containing the three spin matrices as defined in \cite{Kawaguchi2012}
\begin{equation}
    \boldsymbol{f} = (f_x,f_y,f_z)^T \, ,
\end{equation}
and the spin density $\boldsymbol{F}$(x,t) is given by
\begin{equation}
    \boldsymbol{F}=\begin{pmatrix}
        F_x \\
        F_y \\
        F_z
    \end{pmatrix}
    =
    \frac{1}{\sqrt{2}}
    \begin{pmatrix}
        \psi_0^*\left(\psi_{+1}+\psi_{-1}\right)+\mathrm{c.c.}\,\\
        \mathrm{i}\left[\psi_0^*\left(\psi_{+1}-\psi_{-1}\right)+\mathrm{c.c.}\,\right]\\
        \sqrt{2} \left(|\psi_{+1}|^2-|\psi_{-1}|^2\right)
    \end{pmatrix}.
    \label{eq:Spins}
\end{equation}
In the following, we want to numerically investigate effects that are due to the non-perfect experimental state preparation.
The state onto which the spinor phase imprint is made is chosen as either the ground state of the easy-plane ferromagnetic phase at $q=0$, that is
\begin{equation}
    \boldsymbol{\psi}_{\mathrm{EP}, q=0} = \frac{\sqrt{n}}{2}\begin{pmatrix}
        1 \\
        \sqrt{2} \\
        1
    \end{pmatrix} \:,
\end{equation}
corresponding to a fully elongated spin (this is similar to the experimental state preparation), or as the ground state of the easy-plane phase for the chosen $q$,
\begin{equation}
    \boldsymbol{\psi}_\mathrm{EP} = \frac{\sqrt{n}}{2}\begin{pmatrix}
        \sqrt{1-\frac{q}{2n|c_1|}}\\
         \sqrt{2(1+\frac{q}{2n|c_1|}}) \\
        \sqrt{1-\frac{q}{2n|c_1|}}
    \end{pmatrix} \:.
\end{equation}
If one does not use the appropriate ground state for a chosen $q$, the spin background of the system will undergo spin-length oscillations, as can be seen in Fig.~\ref{fig:simulation_kink_q03} for a moving soliton.

\subsection{Experiment-like imprint}
To simulate a kink close to the experimental imprint the initial state is rotated in spin space with the generator of the rotation given by
\begin{equation}
    Q_0 = \begin{pmatrix}
        -1 & 0 & 0 \\
        0 & 1 & 0 \\
        0 & 0 & -1 \\
    \end{pmatrix} \:, 
\end{equation}
with a position-dependent angle
\begin{equation}
    \varphi_\mathrm{rot}(x) = \begin{cases}\frac{\pi}{2} \exp(-\frac{\left(x-x_0\right)^2}{2\xi_\mathrm{s}^2}) \:,& \text{if } x\leq x_0 \:; \\
    \frac{\pi}{2} \:,& \text{if } x > x_0 \:.
    \end{cases}
    \label{eq:kink_simulation}
\end{equation}
This state preparation closely mimics the experimental imprint technique. The standard deviation of the Gaussian edge is chosen as the spin healing length
\begin{equation}
    \xi_\mathrm{s} = \frac{\hbar}{\sqrt{2Mn|c_1|}}\:,
\end{equation}
approximately corresponding to the size of the beam used in the experiment. 
This leads to the initial state:
\begin{equation}
    \boldsymbol{\psi}_\mathrm{init}(x) = \boldsymbol{\psi}_\mathrm{EP}\cdot e^{-i\varphi_\mathrm{rot}(x)Q_0} \; .
\end{equation}

\begin{figure}
    \centering
    \includegraphics[width=0.49\textwidth]{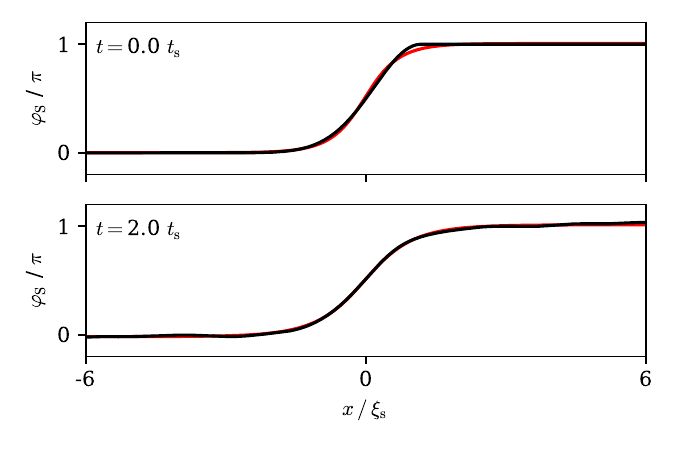}
    \caption{The spinor phase in the simulation (black) for $q=0$ for two times steps with a kink of the shape $a\cdot \arctan{\left(\exp{\left((x-x_0)/\ell\right)}\right)+c}$, similar to Eq.~\eqref{eq:kink_analytics}, fitted (in red) to the data. For $t=0$, the fit gives  $a=2.01$, $x_0=-0.02 \; \xi_\mathrm{s}$, $\ell=0.47 \; \xi_\mathrm{s}$, $c=0$ and for $t=2.0 \; t_\mathrm{s}$ it gives $a=2.06$, $x_0=-0.03 \; \xi_\mathrm{s}$, $\ell=0.67 \; \xi_\mathrm{s}$, $c=-0.05$.}
    \label{fig:simulation_spinorphase_slices}
\end{figure}

How this compares to the analytical solution for kinks ($K$) and anti-kinks ($\overline{K}$) in the sine-Gordon model \cite{Cuevas-Maraver2014},
\begin{equation}
    \varphi_{S,\mathrm{K}/\overline{\mathrm{K}}} = 2\arctan\left[\exp\left(\pm \frac{x-x_0}{\ell}\right)\right]\: ,
    \label{eq:kink_analytics}
\end{equation}
is shown in Fig.~\ref{fig:simulation_spinorphase_slices}. In this analytical solution the width of the (anti-)kink $\ell$ is taken as
\begin{equation}
    \ell = \frac{1}{\sqrt{2}}\xi_\mathrm{s} = \frac{\hbar}{\sqrt{4Mn|c_1|}}\:,
\end{equation}
as given in \cite{Yu2021}.
To obtain the initial state from this phase, one adds $\phis$ as a  complex phase to the $m=0$ component:
\begin{equation}
    \boldsymbol{\psi}_\mathrm{init} = \begin{pmatrix}
        \psi_{\mathrm{EP},+1} \\
        \psi_{\mathrm{EP},0}\cdot e^{-\mathrm{i} \phis} \\
        \psi_{\mathrm{EP},-1}
    \end{pmatrix} \;.
\end{equation}

\subsection{Free propagation}

\begin{figure}
    \centering
    \includegraphics[width=0.5\textwidth]{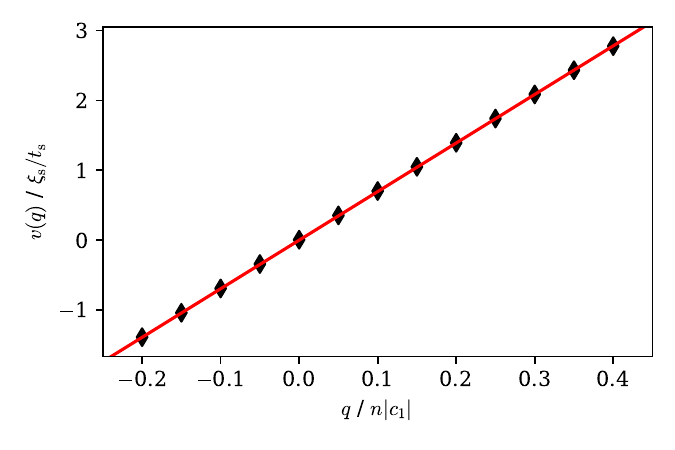}
    \caption{The velocity of the kink for different $q$. The slope extracted from the fit is $\alpha=(49.585\pm 0.005)\, \text{µm}$/(s$\cdot$Hz).}
    \label{fig:v_vs_q_simulation}
\end{figure}

As the simulation uses periodic boundary conditions, we choose a kink-anti-kink pair as initial condition. In the following, we also use the analytical kink profile as the initial condition in the spinor phase. In order to observe a single soliton, the initial distance between the two is chosen as $d_\mathrm{init}>20\,\xi_\mathrm{s}$, such that the solitons do not interact during the whole time evolution.
From these simulations, one can see that the imprint remains stable over time in all relevant observables.
As observed in the experiment, a change in $q$ leads to a change in the velocity of the soliton. When repeating the simulation for different $q$ and using a linear fit to extract the velocities, one finds an approximately linear relation between $v$ and $q$ for the simulated values of $q$ (see Fig. \ref{fig:v_vs_q_simulation}).

\subsection{Collision}
\begin{figure}
    \centering
    \includegraphics[width=0.5\textwidth]{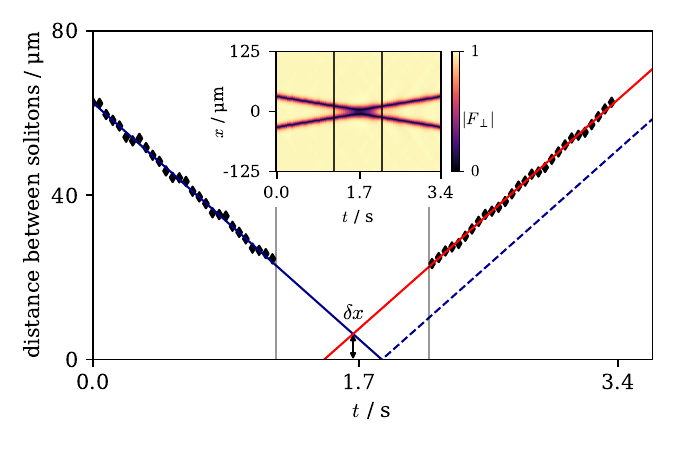}
    \caption{Phase shift extraction from the simulation. Here the incoming velocity is $v_\mathrm{in}=16.82 \, \text{µm/s}$ and the outgoing velocity is $v_\mathrm{out}=16.75 \,\text{µm/s}$.}
    \label{fig:collision_numerics}
\end{figure}

To obtain a numerical value for the phase shift $\delta x$ that can be compared to the value obtained from the experimental results, we choose the density in the simulation such that the width of the soliton in the simulation and the experiment coincide. We perform the same data analysis as in the experiment in each numerical run, which leads to the results shown in Fig.~\ref{fig:collision_numerics}. This gives a phase shift of $\delta x = 6.23 \, \text{µm}$ with an elasticity of $0.996$.

\begin{figure}
    \centering
    \includegraphics[width=0.5\textwidth]{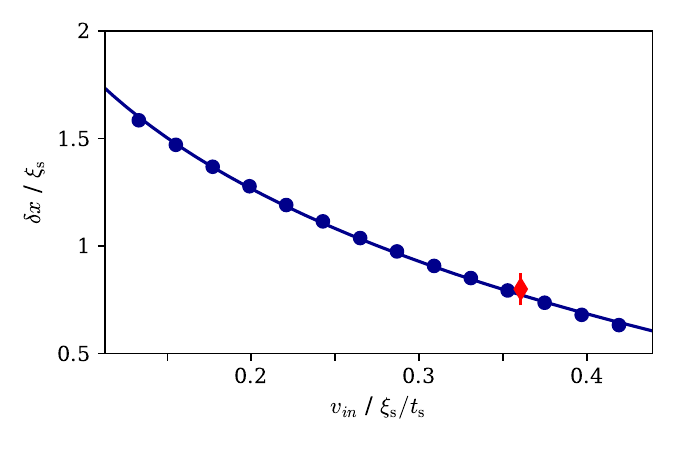}
    \caption{Phase shifts extracted from spin 1 simulations for different velocities. We fit the formula given in Eq.~\eqref{eq:AnalyticsPhaseShift} to the data and obtain an effective mass of $m_{\mathrm{fit}} = 2.533\pm0.019$. The red point is the experimentally extracted phase shift discussed in the main text.}
    \label{fig:numerical_phase_shift}
\end{figure}


\section{Velocity of the soliton}\label{app:Velocity}
In this section, we derive the dependence of the spinor phase-kink velocity on the quadratic Zeeman shift $q$ as given by Eq.~\eqref{eq:velocity} in the main text. To this end, we first derive the spin-nematic continuity equation of the spin-1 Bose gas \cite{Yu2022}. We use this to calculate the velocity dependence of the spinor phase solitons on the quadratic Zeeman shift $q$. 


\subsection{Spin-nematic continuity equation} 
We begin by calculating the time derivative of the magnetization density
\begin{align}
    \partial_t F_i = \partial_t \left( \boldsymbol{\psi}^\dagger f_i \boldsymbol{\psi} \right)
    = (\partial_t \boldsymbol{\psi}^\dagger) f_i \boldsymbol{\psi} + \boldsymbol{\psi}^\dagger f_i (\partial_t \boldsymbol{\psi}),
    \label{eq:spin_derivative}
\end{align}
which is the observable of interest.
Next, we define the differential operator given by the one-dimensional GPE as 
\begin{align}
    \mathcal{D}_\mathrm{GPE} = &-\frac{\hbar^2}{2M}\partial_x^2+V(x)+pf_z+qf_z^2  \\&+\frac{1}{2}c_0|\boldsymbol{\psi}(x,t)|^2+\frac{1}{2}c_1\boldsymbol{F}(x,t)\boldsymbol{f}\, , \nonumber
\end{align}
and write the GPE and its Hermitian conjugate in compact notation,
\begin{align}
    \partial_t \boldsymbol{\psi} = -\frac{\mathrm{i}}{\hbar} \mathcal{D}_\mathrm{GPE}\boldsymbol{\psi} \, , \quad
    \partial_t \boldsymbol{\psi}^\dagger = \frac{\mathrm{i}}{\hbar} \boldsymbol{\psi}^\dagger \mathcal{D}_\mathrm{GPE} \, ,
\end{align}
where, for brevity, the spatial derivatives in $\mathcal{D}_\mathrm{GPE}$ act to the left in the Hermitian conjugate case.
Hence, we can write Eq. \eqref{eq:spin_derivative} as
\begin{align}
    \partial_t F_i = \frac{\mathrm{i}}{\hbar} \boldsymbol{\psi}^{\dagger} \left[\mathcal{D}_\mathrm{GPE}, f_i\right] \boldsymbol{\psi}.
    \label{eq:spin_evolution}
\end{align}
The kinetic term in Eq.~\eqref{eq:spin_evolution} can be written in the form of a spin current term
\begin{align}
    {J}^\mathrm{F}_i = \frac{\hbar}{2M \mathrm{i}}\left[ \boldsymbol{\psi}^{\dagger} f_i \partial_x \boldsymbol{\psi} - (\partial_x \boldsymbol{\psi}^{\dagger}) f_i\boldsymbol{\psi}  \right] \, .
\end{align}
The commutator of the density interaction term in Eq.~\eqref{eq:spin_evolution} vanishes. Therefore, we write the time evolution of the spin density with the remaining terms containing spin matrices as
\begin{align}
     \partial_t F_i &=  -\partial_x {J}^\mathrm{F}_i \nonumber \\ &+ \frac{\mathrm{i}}{\hbar} \boldsymbol{\psi}^{\dagger} \left[pf_z+qf_z^2+\frac{1}{2}c_1\boldsymbol{F}(x,t)\boldsymbol{f}, f_i\right] \boldsymbol{\psi}. 
\end{align}
The next step involves calculating the commutators arising from the spin interactions and the linear and quadratic Zeeman shifts using the $\mathfrak{su}(2)$ commutation relations \\ 
\begin{align}
    \left[\frac{c_1}{2}F_j f_j, f_i\right] &= \mathrm{i} \frac{c_1}{2}F_j  \epsilon_{jik} f_k\, , \\[0.2cm]
    p \, [f_z, f_i] &=  \mathrm{i} \, p \epsilon_{zik} f_k, \, \\[0.25cm]
    q \; [f_zf_z, f_i] &= \mathrm{i} \, q \epsilon_{zik} \{f_z, f_k\}\, , \label{eq:Nematic_qpart}
\end{align}
where $\epsilon_{ijk}$ is the three-dimensional antisymmetric Levi-Civita symbol.
Collecting all terms, we arrive at
\begin{align}
    \partial_t F_i = &-\partial_x {J}^\mathrm{F}_i 
    - \frac{1}{\hbar}\boldsymbol{\psi}^{\dagger}\Big(\frac{c_1}{2}F_j \epsilon_{jik} f_k  \nonumber \\[0.25cm] 
    &\hspace*{1.5cm}+ p \epsilon_{zik} f_k + q\epsilon_{zik}\{f_z, f_k\}\Big)\boldsymbol{\psi}. \label{eq:spin_evolution_anticomm}
\end{align}

The last term in the first line of Eq.~\eqref{eq:spin_evolution_anticomm} vanishes due to the anti-symmetry of the Levi-Civita symbol. The term proportional to $q$ can be written as a nematic source term
$K_{iz} = \sum_k \epsilon_{izk} \boldsymbol{\psi}^{\dagger} Q_{zk} \boldsymbol{\psi}$
in terms of the normalized nematic operator
$Q_{ij} = (f_if_j + f_jf_i)/2$,
ultimately leading to the spin-nematic continuity equation
\begin{align}
    \frac{\partial}{\partial t} F_i + \partial_x {J}_i^\mathrm{F} = \frac{1}{\hbar} \left(2qK_{iz} + p\epsilon_{zik} F_k \right)\, . 
    \label{eq:Spin-Nematic-continuity}
\end{align}


\subsection{Spinor-phase soliton velocity from spin-continuity equation} 
In the following, we derive the dependence of the spinor-phase kink from the spin nematic continuity equation~\eqref{eq:Spin-Nematic-continuity}.
The linear Zeeman shift in the spin-1 Hamiltonian can be absorbed into the fields by considering a rotating frame of reference, hence setting $p=0$. Next, we consider the core of the magnetic soliton, where the magnetization vanishes, leading to a net zero current ${J}_i^\mathrm{F} = 0$. However, its spatial derivative 
$\partial_x{J}_i^\mathrm{F}$
does not necessarily vanish.
We consider a sine-Gordon kink solution in the spinor phase,
\begin{align}
    \psi_{0} = \sqrt{n_{0}}\e^{- \frac{\mathrm{i}}{2} \varphi_{\mathrm{S},\mathrm{K}}},
    \quad \psi_{\pm1} = \sqrt{n_{\pm1}}\e^{\frac{\mathrm{i}}{2} \varphi_{\mathrm{S},\mathrm{K}}},
    \label{eq:hydro_parametrization_spinor}
\end{align}
with
\begin{align}
    \varphi_{\mathrm{S},\mathrm{K}}(x) = 2 \arctan{(\e^{x/\ell})}, \label{eq:BlakieLösung}
\end{align}
and $\ell = \hbar/\sqrt{4|c_1|Mn}$ as in Eq.~\eqref{eq:kink_analytics}. We remark that this class of solutions was shown to solve the spin-1 GPE in the easy-plane phase \cite{Yu2021}. Generally, the following calculation holds for any phase profile, since we evaluate the final solution at the core of the soliton. In particular, it holds for the anti-kink $\overline{K}$ as well, but with a relative factor of $(-1)$ in $\phis$. 
We boost the stationary solution $x \mapsto x - vt$ and obtain a time-dependent solution with velocity $v$. Hence, the time derivative of the phase factors in Eq.~\eqref{eq:hydro_parametrization_spinor} reads
\begin{align}
    \frac{\partial }{\partial t} \e^{\pm \mathrm{i} \varphi_{\mathrm{S}, \mathrm{K}}(x -vt)} &= \mp 2\mathrm{i} \frac{v}{\ell} \frac{\e^{(x-vt)/\ell}}{\e^{2(x-vt)/\ell} + 1} \e^{\pm \mathrm{i} \varphi_{\mathrm{S}, \mathrm{K}}(x-vt)} \, . 
\label{eq:der_arctan}
\end{align}
Notice that we neglect a factor of one half in the exponent, since it is missing in the phase factors of the observable $F_x$ as well.
We evaluate the expression at the core of the soliton, $x-vt=0$, 
which yields
\begin{align}
     \left.\frac{\partial }{\partial t} \e^{\pm \mathrm{i} \varphi_{\mathrm{S}, \mathrm{K}}(x-vt)}\right|_{x=vt}=  \pm \frac{v}{\ell} .
\end{align}
Considering the spin in the $x$ direction, given by Eq.~\eqref{eq:Spins}, we compute the time evolution of the spin, i.e., the first part of the left-hand side of the continuity equation~\eqref{eq:Spin-Nematic-continuity}
\begin{align}
    \frac{\partial}{\partial t} F_x = 
    &\frac{1}{\sqrt{2}} \frac{\partial}{\partial t} \left[\sqrt{n_{-1}n_0}\e^{-\mathrm{i} \varphi_{\mathrm{S},\mathrm{K}}}\right. \nonumber \\ 
    &+ \left.\sqrt{n_0}(\sqrt{n_{+1}}+ \sqrt{n_{-1}})\e^{\mathrm{i} \varphi_{\mathrm{S},\mathrm{K}}} + \sqrt{n_{+1}n_0} \e^{-\mathrm{i} \varphi_{\mathrm{S},\mathrm{K}}} \right] \nonumber \\
    = &\frac{1}{\sqrt{2}}\frac{v}{l} \left[\mathrm{i} \; \psi^*_{-1}\psi_0 -\mathrm{i} \; \psi_0^*(\psi_{-1} + \psi_1) +\mathrm{i} \;\psi^*_{1}\psi_0\right] \\
    = &-\frac{1}{\sqrt{2}} \frac{v}{\ell}\boldsymbol{\psi}^{\dagger}
     \begin{bmatrix}
0 & -\mathrm{i} & 0 \\
\mathrm{i} & 0 & \mathrm{i} \\
0 & -\mathrm{i} & 0
\end{bmatrix}
\boldsymbol{\psi}.
    \label{eq:RHSconteq_filledout}
\end{align}
Calculating the gradient of the current yields
\begin{align}
    \partial_x {J}_x^\mathrm{F} = \frac{\hbar}{2M\mathrm{i}} \left[\boldsymbol{\psi}^{\dagger} f_x \partial_x^2 \boldsymbol{\psi} - (\partial_x^2 \boldsymbol{\psi}^{\dagger}) f_x \boldsymbol{\psi}\right].
    \label{ed:current_dev}
\end{align}
We then proceed with calculating the second spatial derivative of the phase factor of the moving soliton solution as 
\begin{align}
    \partial_x^2 &\e^{\pm i \varphi_{\mathrm{S, K}}} = \frac{1}{\ell^2}\bigg\{\Big(\pm 2\mathrm{i} \frac{\e^{(x-vt)/\ell}}{\e^{2(x-vt)/\ell}+1}\Big)^2 \nonumber \\
    &\pm 2 \mathrm{i} \; \frac{\e^{(x-vt)/\ell}-\e^{3(x-vt)/\ell} }{(\e^{2(x-vt)/\ell}+1)^2}  \bigg\}\e^{\pm \mathrm{i}\varphi_{\mathrm{S,K}}}\;. 
    \label{eq:second_dev}
\end{align}
Again, evaluating Eq. \eqref{eq:second_dev} at the soliton core, $x-vt = 0$, we obtain
\begin{align}
    \left.\partial_x^2 \e^{\pm \frac{\mathrm{i}}{2} \varphi_{\mathrm{S, K}}}\right|_{x=vt} =  \mp \frac{\mathrm{i}}{2\ell^2} .
\end{align}
Plugging this into Eq. \eqref{ed:current_dev} yields 
\begin{align}
   \partial_x {J}_x^\mathrm{F} =& 
     \frac{\hbar}{2M \mathrm{i}} \frac{\mathrm{i}}{2\ell^2} \frac{1}{\sqrt{2}}\bigg(\boldsymbol{\psi}^{\dagger}
     \begin{bmatrix}
0 & 1 & 0 \\
1 & 0 & 1 \\
0 & 1 & 0
\end{bmatrix}
\boldsymbol{\psi} \nonumber \\
&- \boldsymbol{\psi}^{\dagger}
     \begin{bmatrix}
0 & 1 & 0 \\
1 & 0 & 1 \\
0 & 1 & 0
\end{bmatrix}
\boldsymbol{\psi} \bigg) = 0.
\end{align}
Both terms already equate to zero individually, since $F_x$ vanishes at the soliton core and the derivative only introduces a factor. 
The right-hand side of the continuity equation reads
\begin{align}
   \frac{2q}{\hbar} K_{xz} &= \frac{q}{\hbar} \epsilon_{xzy}  \boldsymbol{\psi}^{\dagger} (f_zf_y + f_yf_z) \boldsymbol{\psi} \\
   &= -\frac{q}{\hbar} \frac{1}{\sqrt{2}} \boldsymbol{\psi}^{\dagger}
   \begin{bmatrix}
0 & -\mathrm{i} & 0 \\
\mathrm{i} & 0 & \mathrm{i} \\
0 & -\mathrm{i} & 0
\end{bmatrix}
\boldsymbol{\psi}  \,. \label{eq:LHSconteq_filledout}
\end{align}
Plugging all expressions into Eq. \eqref{eq:Spin-Nematic-continuity}, 
we arrive at a linear relation,
\begin{align}
    \frac{v}{\ell} = \frac{q}{\hbar} \; ,
\end{align}
which recovers Eq.~\eqref{eq:velocity} in the main text.
Note that a density dependence enters via the natural size of the soliton solution, which is proportional to the spin healing length, i.e.  $\ell^{-2}\sim n$.

\subsection{Spinor-phase soliton velocity from effective sG theory\label{VelocityCalcEFT}}
In the following, we give an alternative derivation of Eq.~\eqref{eq:velocity} within the effective sG theory.
With the Lagrangian given in \cite{Siovitz2025} (where only density fluctuations for small momenta were taken into account in the derivation of the effective theory)
the double sine-Gordon equation is found to be:
\begin{align}
         0= \; &\partial^2_{\overline{t}}\phis-2(1-\overline{q}^2)\partial^2_{\overline{x}}\phis +2(1-2\overline{q}^2)\sin{(2\phis)}\nonumber \\ &+\overline{q}^2\sin{(4\phis)} \; .
\end{align}
Here $\overline{x}$ and $\overline{t}$ are the space and time coordinates in units of spin healing length and spin interaction time, and $\overline{q}$ is $q$ in units of $2n|c_1|$.
We rescale $\overline{x}$ as $\Tilde{x} = \overline{x}/\sqrt{2(1-\overline{q}^2)}$.
In the sine-Gordon limit, the solution for a single kink or anti-kink with dimensionless velocity $\overline{v}$ and offset $\Tilde{x}_0$ for this equation is:
\begin{equation}
    \phis=\pm2\arctan{\left(\exp{\sqrt{2}\sqrt{\frac{1-2\overline{q}^2}{1-\overline{q}^2}}\frac{\Tilde{x}-\overline{v} \overline{t}-\Tilde{x}_0}{\sqrt{1-\overline{v}^2}}}\right)}
\end{equation}
As we find the soliton to have a constant width of $\ell= \xi_\mathrm{s}/\sqrt{2}$, which in units of $\overline{x}$ is $\overline{\ell}=1/\sqrt{2}$, the factors in the argument of the exponential function should multiply to the inverse of $\overline{\ell}$, leading to:
\begin{equation}
    \frac{1}{\overline{\ell}} = \sqrt{2} = \sqrt{2}\sqrt{\frac{1-2\overline{q}^2}{1-\overline{q}^2}}\frac{1}{\sqrt{1-\overline{v}^2}} \; .
\end{equation}
When solving this for the dimensionless velocity, one obtains:
\begin{equation}
    \overline{v}=\frac{\overline{q}}{\sqrt{1-\overline{q}^2}} \; .
\end{equation}
This can be transformed back to the observed velocity $v_{\mathrm{obs}}$ in physical units:
\begin{align}
    v_{\mathrm{obs}}= \; &\frac{\overline{t}}{t}\frac{x}{\overline{x}}\sqrt{2(1-\overline{q}^2)}\overline{v} \nonumber \\
    = \; &\sqrt{2}\frac{\overline{t}}{t}\frac{x}{\overline{x}}\overline{q} \nonumber \\
    = \; & \frac{\xi_\mathrm{s}}{\sqrt{2}\hbar}q \; .
\end{align}
With $\ell=\xi_\mathrm{s}/\sqrt{2}$, this gives:
\begin{equation}
    \frac{v_{\mathrm{obs}}}{\ell} = \frac{q}{\hbar} \; .
\end{equation}
We are therefore able to derive the velocity-$q$ dependence even in the effective sG theory, thereby further showing that the experimentally generated spinor-phase kink is described by an effective sG theory.

\section{Phase shift from sG theory}\label{app:PhaseShift}

In this section, we calculate the value of the phase shift from the effective sG theory derived in \cite{Siovitz2025} in order to see if the effective theory is also able to quantitatively describe soliton collisions. We start from the equation of motion
\begin{equation}
    \partial_{\bar{t}}^2 \phis - 2(1-\bar{q}^2) \partial_{\bar{x}}^2 \phis + 4(1-2 \bar{q}^2) \mathrm{sin}(\phis) = 0 \; .
\end{equation}
To obtain an analytical estimate of the phase shift $\delta x$, we use the formula for the phase shift in the sine-Gordon model in standard form,
\begin{equation}\label{eq:AnalyticsPhaseShift}
    \delta x = \frac{2 \sqrt{1-(v/c_\mathrm{s})^2}}{m}\ln{\frac{c_\mathrm{s}}{v}},
\end{equation}
as calculated in \cite{Koch2023}, where $m=\sqrt{\lambda}$ is the mass of the sine-Gordon field.
The mass of the effective theory derived in \cite{Siovitz2025}  is given by
$m=2\sqrt{1-2\overline{q}^2}$, with $\overline{q}=q/2n|c_1|$, and an additional factor in the transformation of space to dimensionless coordinates of $1 / \sqrt{2(1-\overline{q}^2)}$.
This leads to the following relations for the phase shift $\delta\overline{x}$ in units of $\overline{x}$:
\begin{align}
    \delta\overline{x}&= \frac{\sqrt{2-2\overline{q}^2 - \overline{v}^2}}{\sqrt{1-2\overline{q}^2}}\ln{\frac{\sqrt{2(1-\overline{q}^2)}}{\overline{v}}}\:,\quad 
\end{align}
To obtain the velocity $\overline{v}$ in units of $\overline{x}$ and $\overline{t}$ and then also the phase shift $\delta x$ in physical units, the relations $x=\xi_\mathrm{s}\overline{x}=\sqrt{2}\ell \overline{x}$ and $t=\frac{2M\xi_\mathrm{s}^2}{\hbar}\overline{t}=\frac{4M\ell^2}{\hbar}\overline{t}$ \cite{Siovitz2025} are used.
The width of the collision data in the experiment was found as the mean of the fit results of the soliton widths as $\ell=5.57 \;\text{µm}$.
With an incoming velocity obtained from the fit as $v_{\mathrm{in}}=16.76\; \text{µm}/\text{s}$ while keeping the aforementioned approximation $\bar{q}=0$, this gives $\delta x = 14.30 \;\text{µm}$. \\
This value is more than twice as large as the measured (and numerical) value for the phase shift. This is, however, not surprising as the assumption of constant densities $n_{\mf}$ made in the derivation of the effective sG theory in \cite{Siovitz2025} is violated during collision. We therefore have fitted Eq.~\eqref{eq:AnalyticsPhaseShift} to phase shifts extracted from numerical spin-1 simulation, see Fig.~\ref{fig:numerical_phase_shift},  and found that with a different sG mass of $m_{\mathrm{fit}} = 2.53\pm0.02$ we can still capture the behavior during collision in an effective sG theory.

\end{appendix}
\twocolumngrid
\bibliographystyle{apsrev4-1}

\bibliography{Bibliography}

\end{document}